\documentclass{article}
\usepackage{latexsym}
\usepackage{graphicx}
\begin{document}
\begin{center}
\textbf{\LARGE Potential Theory in Classical Electrodynamics}
\end{center}
\begin{large}
\begin{center}
W. Engelhardt\footnote{Home address: Fasaneriestrasse 8, D-80636 M\"{u}nchen, Germany\par \hspace*{.15cm} Electronic address: wolfgangw.engelhardt@t-online.de}, retired from:
\end{center}

\begin{center}
Max-Planck-Institut f\"{u}r Plasmaphysik, D-85741 Garching, Germany
\end{center}

\vspace{.6cm}

\noindent \textbf{Abstract}

\noindent In Maxwell's classical theory of electrodynamics the fields are frequently expressed by potentials in order to facilitate the solution of the first order system of equations. This method obscures, however, that there exists an inconsistency between Faraday's law of induction and Maxwell's flux law. As a consequence of this internal contradiction there is neither gauge invariance, nor exist unique solutions in general. It is also demonstrated that inhomogeneous wave equations cannot be solved by retarded integrals.
\vspace{.6cm}

\vspace{1.cm}

\noindent \textbf{1 Introduction}

\noindent Maxwell's first order system of equations \textit{in vacuo} specifies the divergence and the curl of the electromagnetic fields $\vec {E}$ and$\vec {B}$. In principle, this allows to calculate the fields from the given sources $\rho $ and$\vec {j}$, but the equations are coupled which confronts us with certain complications. In particular, it is not guaranteed that a unique solution exists at all, as it was questioned in [1]. The two source-free equations impose a necessary condition on the fields, if they exist: They must be derivable from a vector and a scalar potential in the following way [2]:
\begin{equation}
\label{eq1}
\vec {B}=\nabla \times \vec {A}\,,\quad \vec {E}=-\nabla \phi -\frac{1}{c}\frac{\partial \vec {A}}{\partial t}
\end{equation}
The potentials itself are to be determined from the inhomogeneous equations which represent a coupled system of second order equations depending on the sources.

Equation (\ref{eq1}) leaves the fields unchanged when a ``gauge transformation'' is imposed on the potentials:
\begin{equation}
\label{eq2}
\vec {A}\to \vec {A}+\nabla \psi \,,\quad \phi \to \phi -\frac{1}{c}\frac{\partial \psi }{\partial t}
\end{equation}
where $\psi $ is an arbitrary differentiable function. As a consequence the divergence of the vector potential is arbitrary to the same extent as the Laplacian $\Delta \psi $. This fact has been exploited to decouple the second order system. There is, however, no proof in the literature that the chosen procedure is viable and leads to unique solutions. In principle, the potentials are defined by the inhomogeneous equations which depend on the divergence of the vector potential $\nabla \cdot \vec {A}$. Whether this dependence cancels when the solutions of the second order system are substituted into (\ref{eq1}) is an open, non-trivial question. In Sect. 2 this problem is investigated and it is found that $\nabla \cdot \vec {A}$ cannot be chosen arbitrarily. This explains why solutions for the fields in Lorenz gauge are at variance with those obtained in Coulomb gauge as was found in [1] and [3].

The standard procedure of solving Maxwell's equations in Lorenz gauge leads to decoupled inhomogeneous wave equations for the potentials which are thought to be solved by retarded integrals. In [1], however, it was claimed that the inhomogeneous wave equations cannot be solved in general, since they connect sources and potentials at the same time, whereas in the retarded solutions the potentials and the sources are to be evaluated at different times. This ambiguity is again analyzed in Sect. 3 where it is shown for charges moving at constant velocity that the Li\'{e}nard-Wiechert scalar potential cannot be considered as a solution of the inhomogeneous wave equation.

It turns out then that the system of Maxwell's first order equations does not permit a solution in general. Only the homogeneous wave equations, which were exclusively considered by Maxwell in the context of his theory of light [1], are suitable to describe travelling electromagnetic waves which are disconnected from their sources. The problem is deeply rooted in an inconsistency of the first order system which is usually concealed by the potential ansatz (\ref{eq1}). Analyzing the fields inside a plate capacitor the ambiguity is made visible in Sect. 4. Concluding remarks in Sect. 5 terminate this study on the potential method. 

\vspace{1. cm} 
\noindent \textbf{2 Dependence of the fields on the divergence of the vector potential}

\noindent When we substitute the potential ansatz (\ref{eq1}) into the inhomogeneous Maxwell equations we obtain the system: 
\begin{equation}
\label{eq3}
\Delta \phi =-4\pi \,\rho -\frac{1}{c}\frac{\partial \chi }{\partial t}
\end{equation}
\begin{equation}
\label{eq4}
\Delta \vec {A}-\frac{1}{c^2}\frac{\partial ^2\vec {A}}{\partial t^2}=-\frac{4\pi }{c}\,\vec {j}+\nabla \chi +\frac{1}{c}\nabla \frac{\partial \phi }{\partial t}
\end{equation}
where the abbreviation $\nabla \cdot \vec {A}=\chi $ was used. Naturally, the gauge function $\psi $ does not enter into these equations, as it cancels according to (\ref{eq1}) in the expressions for the fields. The potentials, however, will become a function of $\chi $ according to (\ref{eq3}) and (\ref{eq4}). One must now investigate whether $\chi $ will also cancel in (\ref{eq1}), when the solutions of (\ref{eq3}) and (\ref{eq4}) are substituted. To this end one can exploit the linearity of eqs. (3, 4) and split them in the following way:
\begin{equation}
\label{eq5}
\phi =\phi _1 +\phi _2 
\end{equation}
\begin{equation}
\label{eq6}
\Delta \phi _1 =-4\pi \,\rho \;
\end{equation}
\begin{equation}
\label{eq7}
\Delta \phi _2 =-\frac{1}{c}\frac{\partial \chi }{\partial t}\;
\end{equation}
This set of equations is entirely equivalent to (\ref{eq3}). Similarly:
\begin{equation}
\label{eq8}
\vec {A}=\vec {A}_1 +\vec {A}_2 
\end{equation}
\begin{equation}
\label{eq9}
\Delta \vec {A}_1 -\frac{1}{c^2}\frac{\partial ^2\vec {A}_1 }{\partial t^2}=-\frac{4\pi }{c}\,\vec {j}+\frac{1}{c}\nabla \frac{\partial \phi _1 }{\partial t}\;
\end{equation}
\begin{equation}
\label{eq10}
\Delta \vec {A}_2 -\frac{1}{c^2}\frac{\partial ^2\vec {A}_2 }{\partial t^2}=\nabla \chi +\frac{1}{c}\nabla \frac{\partial \phi _2 }{\partial t}\;
\end{equation}
Applying Helmholtz's theorem on the vector potential Chubykalo et al. [4] have shown that the set of equations (\ref{eq6}) and (\ref{eq9}) determines the fields uniquely, when the solutions $\phi _1 $ and $\vec {A}_1 $ are substituted into (\ref{eq1}). In a comment by V. Onoochin and the present author [5] it was pointed out that Chubykalo's procedure is equivalent to choosing $\chi =0$, or adopting Coulomb gauge. It follows then by insertion of (\ref{eq5}) and (\ref{eq8}) into (\ref{eq1})
\begin{equation}
\label{eq11}
\vec {B}=\nabla \times \vec {A}_1 +\nabla \times \vec {A}_2 \,,\quad \vec {E}=-\nabla \phi _1 -\frac{1}{c}\frac{\partial \vec {A}_1 }{\partial t}-\nabla \phi _2 -\frac{1}{c}\frac{\partial \vec {A}_2 }{\partial t}
\end{equation}
that the terms containing $\chi $ must vanish separately
\begin{equation}
\label{eq12}
\nabla \times \vec {A}_2 =0
\end{equation}
\begin{equation}
\label{eq13}
\nabla \phi _2 +\frac{1}{c}\frac{\partial \vec {A}_2 }{\partial t}=0
\end{equation}
in order to render the fields independent of the chosen gauge $\chi$. 

Let us check whether the solutions of (\ref{eq7}) and (\ref{eq10}) for an arbitrary choice of $\chi $ satisfy the conditions (\ref{eq12}) and (\ref{eq13}). First we notice that $\vec {A}_2 $ must satisfy the necessary condition 
\begin{equation}
\label{eq14}
\vec {A}_2 =\nabla U
\end{equation}
because of (\ref{eq12}). Inserting this into (\ref{eq13}) yields
\begin{equation}
\label{eq15}
\phi _2 +\frac{1}{c}\frac{\partial U}{\partial t}=0
\end{equation}
and equation (\ref{eq10}) becomes
\begin{equation}
\label{eq16}
\Delta U-\frac{1}{c^2}\frac{\partial ^2U}{\partial t^2}=\chi +\frac{1}{c}\frac{\partial \phi _2 }{\partial t}\;
\end{equation}
The retarded solution of this wave equation -- subject to the boundary condition $U\left( {\vec {\infty }} \right)=0$ -- is:
\begin{equation}
\label{eq17}
U=\frac{-1}{4\pi \,}\int\!\!\!\int\!\!\!\int_V {\frac{d^3x'}{\left| {\vec {x}-\vec {x}\,'} \right|}} \left[ {\chi \left( {\vec {x}\,',\,t'} \right)+\frac{1}{c}\frac{\partial \phi _2 \left( {\vec {x}\,',\,t'} \right)}{\partial t'}} \right]_{t'=t-{\left| {\vec {x}-\vec {x}\,'} \right|} \mathord{\left/ {\vphantom {{\left| {\vec {x}-\vec {x}\,'} \right|} c}} \right. \kern-\nulldelimiterspace} c} 
\end{equation}
The instantaneous solution of the Poisson equation (\ref{eq7}) under the boundary condition $\phi _2 \left( {\vec {\infty }} \right)=0$ is:
\begin{equation}
\label{eq18}
\phi _2 =\frac{1}{4\pi \,c}\int\!\!\!\int\!\!\!\int_V {\frac{d^3x'}{\left| {\vec {x}-\vec {x}\,'} \right|}} \frac{\partial \chi \left( {\vec {x}\,',\,t} \right)}{\partial t}
\end{equation}
where the integration has to be carried out over all space. Substituting (\ref{eq18}) into (\ref{eq15}) yields an instantaneous solution for $U$ after integration with respect to time:
\begin{equation}
\label{eq19}
U=\frac{-1}{4\pi \,}\int\!\!\!\int\!\!\!\int_V {\frac{d^3x'}{\left| {\vec {x}-\vec {x}\,'} \right|}} \,\chi \left( {\vec {x}\,',\,t} \right)
\end{equation}
that is not compatible with (\ref{eq17}) for an arbitrary function $\chi \left( {\vec {x},\,t} \right)$. Choosing, for example,
\begin{equation}
\label{eq20}
\chi =\frac{4}{\sqrt \pi \,d^3}\,\exp \left( {-{r^2} \mathord{\left/ {\vphantom {{r^2} {d^2}}} \right. \kern-\nulldelimiterspace} {d^2}} \right)\,\sin \omega \,t\;,\quad r=\sqrt {x^2+y^2+z^2} 
\end{equation}
equation (\ref{eq18}) yields:
\begin{equation}
\label{eq21}
\phi _2 =\frac{\omega }{c}\frac{\mbox{erf}\left( {r \mathord{\left/ {\vphantom {r d}} \right. \kern-\nulldelimiterspace} d} \right)}{r}\,\cos \omega \,t
\end{equation}
and (\ref{eq19}) results in:
\begin{equation}
\label{eq22}
U=-\frac{\mbox{erf}\left( {r \mathord{\left/ {\vphantom {r d}} \right. \kern-\nulldelimiterspace} d} \right)}{r}\,\sin \omega \,t
\end{equation}
On the other hand, one has from (\ref{eq17}) and (\ref{eq21}) the result
\begin{equation}
\label{eq23}
U=-\,\int\!\!\!\int\!\!\!\int_V {\frac{d^3x'}{\left| {\vec {x}-\vec {x}\,'} \right|}} \left[ {\frac{\exp \left( {-{r^2} \mathord{\left/ {\vphantom {{r^2} {d^2}}} \right. \kern-\nulldelimiterspace} {d^2}} \right)}{\pi ^{\frac{3}{2}}\,d^3}-\frac{\omega ^2}{4\pi \,c^2}\frac{erf\left( {r \mathord{\left/ {\vphantom {r d}} \right. \kern-\nulldelimiterspace} d} \right)}{r}} \right]\,\sin \omega \left( {t-{\left| {\vec {x}-\vec {x}\,'} \right|} \mathord{\left/ {\vphantom {{\left| {\vec {x}-\vec {x}\,'} \right|} c}} \right. \kern-\nulldelimiterspace} c} \right)
\end{equation}
The first term may be integrated analytically, but this is not possible for the second one. Obviously, there is a discrepancy between (\ref{eq22}) and (\ref{eq23}) which proves that the necessary and sufficient condition (\ref{eq15}) cannot be met by the solutions (\ref{eq17}) and (\ref{eq18}). Consequently, the electric field expressed by the potentials is a function of $\chi $ in general, as the divergence of the vector potential does not cancel in (\ref{eq1}). 

For the magnetic field one can draw a similar conclusion by writing the inhomogeneous flux equation in integral form:
\begin{equation}
\label{eq24}
\oint {\vec {B}\cdot d\vec {l}} =\frac{1}{c}\mathop{{\int\!\!\!\!\!\int}\mkern-21mu \bigcirc} {\left( {4\pi \,\vec {j}+\frac{\partial \vec {E}}{\partial t}} \right)\cdot d\vec {S}} 
\end{equation}
If $\vec {E}$ depends on $\chi $, this holds also for $\vec {B}$ due to the connection in (\ref{eq24}). On the other hand, if one takes the curl of (\ref{eq10}), one obtains a homogeneous wave equation for $\nabla \times \vec {A}_2 $ which has only the solution $\nabla \times \vec {A}_2 =0$ assuming $\vec {A}_2 \left( {\vec {\infty }} \right)=0$. This implies that the magnetic field does not depend on $\chi $ in agreement with (\ref{eq12}), but in contrast to (\ref{eq24}). In Sect. 4 this ambiguity will be investigated in order to clarify whether Maxwell's equations have unique solutions at all. Before, however, let us analyze an inhomogeneous wave equation of type (\ref{eq4}) and demonstrate in the next Section that it cannot be solved by a retarded integral in general.

\vspace{1.0 cm}
\noindent \textbf{3 Attempt to solve an inhomogeneous wave equation}

\noindent Although the Coulomb gauge $\chi =0$ is the natural gauge, since it follows also from Helmholtz's theorem applied on the vector potential [4], most textbooks make use of the Lorenz gauge 
\begin{equation}
\label{eq25}
\chi =-\frac{1}{c}\frac{\partial \phi _L }{\partial t}
\end{equation}
which results in an inhomogeneous wave equation for the scalar Lorenz potential by substitution into (\ref{eq3}):
\begin{equation}
\label{eq26}
\Delta \phi _L \left( {\vec {x}{\kern 1pt},\,t} \right)-\frac{1}{c^2}\frac{\partial ^2\phi _L \left( {\vec {x}{\kern 1pt},\,t} \right)}{\partial t^2}=-4\pi \,\rho \left( {\vec {x}{\kern 1pt},\,t} \right)
\end{equation}
We may also consider the Poisson equation in Coulomb gauge
\begin{equation}
\label{eq27}
\Delta \phi _C \left( {\vec {x}{\kern 1pt},\,t} \right)=-4\pi \,\rho \left( {\vec {x}{\kern 1pt},\,t} \right)
\end{equation}
and subtract it from (\ref{eq26}) 
\begin{equation}
\label{eq28}
\Delta \left( {\phi _L \left( {\vec {x}{\kern 1pt},\,t} \right)-\phi _C \left( {\vec {x}{\kern 1pt},\,t} \right)} \right)=\frac{1}{c^2}\frac{\partial ^2\phi _L \left( {\vec {x}{\kern 1pt},\,t} \right)}{\partial t^2}
\end{equation}
Note that the charge density cancels, since it is taken both in the instantaneous equation (\ref{eq27}) and in the wave equation (\ref{eq26}) at the same time $t$ when the potentials are evaluated. The formal unique solution for the difference of the potentials is the integral:
\begin{equation}
\label{eq29}
\phi _L \left( {\vec {x}{\kern 1pt},\,t} \right)-\phi _C \left( {\vec {x}{\kern 1pt},\,t} \right)=\int\!\!\!\int\!\!\!\int {d^3x'\,\frac{-1}{\left| {\vec {x}{\kern 1pt}-\vec {x}{\kern 1pt}'{\kern 1pt}} \right|}\left[ {\frac{1}{4\pi \,c^2}\frac{\partial ^2\phi _L \left( {\vec {x}\,'{\kern 1pt},\,t} \right)}{\partial t^2}} \right]} 
\end{equation}
assuming $\phi _C \left( {\vec {\infty }} \right)=\phi _L \left( {\vec {\infty }} \right)=0$. This equation must be satisfied when the individual solutions of (\ref{eq26}) and (\ref{eq27}) are substituted. In order to check on this let us consider a point charge $e$ moving along the $x$-axis with constant velocity $v$. The instantaneous Coulomb potential resulting from (\ref{eq27}) is the well known expression
\begin{equation}
\label{eq30}
\phi _C \left( {x{\kern 1pt},\,t} \right)=\frac{e}{\sqrt {\left( {x-x_0 -\,v\,t} \right)^2+y^2+z^2} }
\end{equation}
where $x_0 $ ist the position of the charge at $t=0$. Li\'{e}nard-Wiechert have calculated the retarded integral for this situation obtaining [1]:
\begin{equation}
\label{eq31}
\phi _L \left( {x{\kern 1pt},\,t} \right)=\frac{e}{\sqrt {\left( {x-x_0 -\,v\,t} \right)^2+\left( {1-{v^2} \mathord{\left/ {\vphantom {{v^2} {c^2}}} \right. \kern-\nulldelimiterspace} {c^2}} \right)\left( {y^2+z^2} \right)} }
\end{equation}
On the $x$-axis expressions (\ref{eq30}) and (\ref{eq31}) are identical so that the integral on the r.h.s. of (\ref{eq29}) must vanish there. This is, however, not the case, if one substitutes solution (\ref{eq31}) into (\ref{eq29}) and carries out the integration over all space. One obtains (see Appendix):
\begin{equation}
\label{eq32}
\phi _C \left( {x{\kern 1pt},\,t} \right)-\phi _L \left( {x{\kern 1pt},\,t} \right)=\frac{e\,v^2}{x\,\left( {c^2-v^2} \right)}
\end{equation}
This proves that the retarded integrals are not proper solutions of the inhomogeneous wave equations which appear not to have a solution at all. The same conclusion was reached in [1] and [6].

\vspace{1.0cm}
\noindent \textbf{4 An inconsistency in determining the magnetic field}

\noindent In order to facilitate the analysis of the flux law (\ref{eq24}) let us consider an axisymmetric case where a plate capacitor is charged up by a variable current (Fig. 1). In cylindrical coordinates one has for the $Z$- component of (\ref{eq24}):
\begin{equation}
\label{eq33}
\frac{1}{R}\frac{\partial \left( {R\,B_\varphi } \right)\,}{\partial R}=\frac{4\pi }{c}\,j_Z +\frac{1}{c}\frac{\partial E_Z }{\partial t}
\end{equation}
In the region between the plates, where the conduction current vanishes, one may integrate (\ref{eq33}) and obtain for the circular magnetic field component
\begin{equation}
\label{eq34}
B_\varphi =\frac{1}{c\,R}\int\limits_0^R {\frac{\partial E_Z }{\partial t}\,} R'dR'
\end{equation}
The quasi-static electric gradient field, which is created by the surface charges on the capacitor plates according to (\ref{eq6}), is easily obtained from the global equations describing a capacitor. One has 
\begin{equation}
\label{eq35}
Q=C\,V\;,\quad I={dQ} \mathord{\left/ {\vphantom {{dQ} {dt}}} \right. \kern-\nulldelimiterspace} {dt}\;,\quad E_Z =V \mathord{\left/ {\vphantom {V d}} \right. \kern-\nulldelimiterspace} d
\end{equation}
where $Q$ is the total charge, $C$ the capacitance, $V$ the voltage, $I$ the current, and $d$ the distance between the plates. Inserting this into (\ref{eq34}) one obtains
\begin{equation}
\label{eq36}
B_\varphi =\frac{1}{c\,R}\int\limits_0^R {\frac{I}{d\,C}\,} R'dR'=\frac{I\,R}{2c\,d\,C}
\end{equation}
for the magnetic field between the plates. In fact, a measurement of this field created by the ``displacement'' current was reported in [7] in agreement with Stokes' law (\ref{eq36}). The slope of this magnetic field was constant according to the results in Fig. 4 of Ref. [7]. Taking the curl of this field one obtains a spatially constant displacement current between the plates that is proportional to the time derivative of $E_Z $ in agreement with (\ref{eq35}). 

\begin{figure}[htbp]
\centerline{\includegraphics[width=6in,height=2.5in]{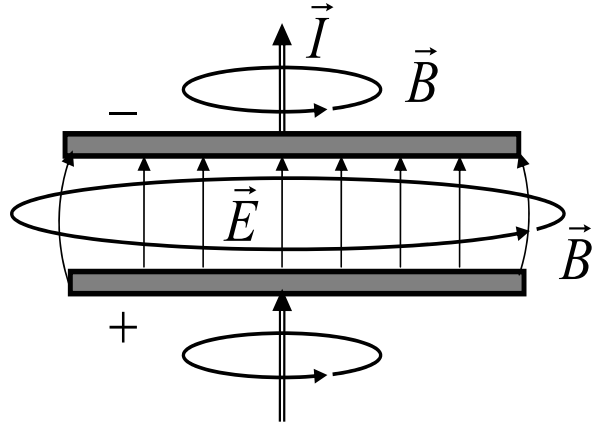}}
\caption{\large Magnetic field created by a quasi-stationary electric field in a plate capacitor}
\label{fig1}
\end{figure}

On the other hand, Faraday's law predicts for the same field component the expression
\begin{equation}
\label{eq37}
\frac{1}{c}\frac{\partial B_\varphi }{\partial t}=\frac{\partial E_z }{\partial R}-\frac{\partial E_R }{\partial z}=0
\end{equation}
as $E_R =0\,,\;E_Z =V \mathord{\left/ {\vphantom {V d}} \right. \kern-\nulldelimiterspace} d$. Equation (\ref{eq37}) yields upon integration over time a magnetic field exclusively created by the solenoidal part of the electric field, whereas the irrotational part in (\ref{eq35}) produced by the charges on the plates does not contribute. According to (\ref{eq37}) the magnetic field between the plates could never change, but in agreement with (\ref{eq36}) and the measurement as reported in [7] the changing electric gradient field is well capable of producing a magnetic field in contrast to the prediction of (\ref{eq37}).

There is apparently an intrinsic inconsistency between Faraday's law of induction and Maxwell's flux law. The discrepancy is not obvious as long as the potential ansatz (\ref{eq1}) is adopted. It guarantees that Faraday's law is satisfied automatically once the vector potential is determined from the flux law. A temporal evolution of the magnetic field, however, is only possible in the presence of a solenoidal electric field $\vec {E}_s $ according to Faraday's law:
\begin{equation}
\label{eq38}
\vec {B}=-c\int {dt} \,\nabla \times \vec {E}_s 
\end{equation}
whereas by spatial integration one finds that the magnetic field is also a function of the changing irrotational electric field $\vec {E}_i $ produced by charge separation:
\begin{equation}
\label{eq39}
\vec {B}=\frac{1}{c}\int\!\!\!\int\!\!\!\int_V {d^3x'} \left( {4\pi \,\vec {j}+\frac{\partial \left( {\vec {E}_s +\vec {E}_i } \right)}{\partial t}} \right)\times \frac{\vec {x}-\vec {x}\,'}{\left| {\vec {x}-\vec {x}\,'} \right|^3}
\end{equation}
In general, equations (\ref{eq38}) and (\ref{eq39}) are incompatible as demonstrated by the discrepancy between (\ref{eq36}) and (\ref{eq37}).

\vspace{1.0cm}
\noindent \textbf{5 Concluding remarks}

\noindent The analysis presented in this paper forces us to recognize that Maxwell's system of first order equations cannot be solved consistently as it contains an internal contradiction. The potential method -- which was already adopted by Maxwell himself -- conceals this fact, but it allows deriving a second order wave equation for the vector potential modelling electromagnetic waves successfully. Maxwell considered the homogeneous wave equation in a region far away from the sources and formulated boundary conditions for its solution [1]. Possibly, he was aware that an inhomogeneous wave equation is not solvable as shown in Sect. 3. This is also true for the vector wave equation (\ref{eq4}) regardless which gauge is chosen.

Classical electrodynamics requires apparently a thorough revision with special attention to the interaction of waves with matter. At this point a concrete proposal is not available, but it is likely that Planck's constant must be built into the system of equations, since it plays a major role in the quantum theory of light that has replaced Maxwell's theory of light.

\vspace{1.0cm}
\noindent \textbf{Appendix}

\noindent In Sect. 2 the volume integral (\ref{eq29}) had to be evaluated:
\begin{eqnarray}
 I=-\frac{1}{4\pi \,c^2}\int\!\!\!\int\!\!\!\int {d^3x'\,\frac{1}{\left| {\vec {x}{\kern 1pt}-\vec {x}{\kern 1pt}'{\kern 1pt}} \right|}\left[ {\frac{\partial ^2\phi _L \left( {\vec {x}\,'{\kern 1pt},\,t} \right)}{\partial t^2}} \right]} \quad\quad\quad\quad\quad\quad\quad\quad \nonumber \\ 
 =-\frac{1}{4\pi \,c^2}\int\!\!\!\int\!\!\!\int {d^3x'\,\frac{1}{\left| {\vec {x}{\kern 1pt}-\vec {x}{\kern 1pt}'{\kern 1pt}} \right|}\frac{\partial ^2}{\partial t^2}\left[ {\frac{e}{\sqrt {\left( {x'-x_0 -v\,t} \right)^2+\left( {1-\beta ^2} \right)\left( {y'^2+z'^2} \right)} }} \right]} \nonumber \\ 
 =\frac{e\,\beta ^2}{4\pi \,}\int\!\!\!\int\!\!\!\int {d^3x'\,\frac{1}{\left| {\vec {x}{\kern 1pt}-\vec {x}{\kern 1pt}'{\kern 1pt}} \right|}\left[ {\frac{-2\,\left( {x'-x_0 -v\,t} \right)^2+\left( {1-\beta ^2} \right)\left( {y'^2+z'^2} \right)}{\left[ {\left( {x'-x_0 -v\,t} \right)^2+\left( {1-\beta ^2} \right)\left( {y'^2+z'^2} \right)} \right]^{\frac{5}{2}}}} \right]} \quad\quad \nonumber \\ \nonumber 
\end{eqnarray}
where $\beta =v \mathord{\left/ {\vphantom {v c}} \right. \kern-\nulldelimiterspace} c$. Changing variables one may write
\[
\vec {x}-\vec {x}\,'=\vec {R}
\]
Evaluation at the time $t=-{x_0 } \mathord{\left/ {\vphantom {{x_0 } v}} \right. \kern-\nulldelimiterspace} v$ when the charge has reached the origin yields on the $x$-axis:
\[
I=\frac{e\,\beta ^2}{4\pi \,}\int\!\!\!\int\!\!\!\int {\,\frac{d^3x'}{R}\left[ {\frac{-2\,\left( {x+R_x } \right)^2+\left( {1-\beta ^2} \right)\left( {R_y^2 +R_z^2 } \right)}{\left[ {\left( {x+R_x } \right)^2+\left( {1-\beta ^2} \right)\left( {R_y^2 +R_z^2 } \right)} \right]^{\frac{5}{2}}}} \right]} 
\]
In spherical coordinates one has
\[
R_x =R\cos \theta \;,\quad R_y =R\cos \varphi \;\sin \theta \;,\quad R_z =R\sin \varphi \;\sin \theta 
\]
and the volume element becomes
\[
d^3x'=R^2\sin \theta \;d\varphi \;d\theta \;dR
\]
This results in:
\[
I=\frac{e\,\beta ^2}{4\pi \,}\int\limits_0^\infty {R\,dR\,\int\limits_0^\pi {\sin \theta \;d\theta \;\int\limits_0^{2\pi } {d\varphi \;} } } \left[ {\frac{-2\,\left( {x+R\,\cos \theta } \right)^2+\left( {1-\beta ^2} \right)\,R^2\,\sin ^2\theta }{\left[ {\left( {x+R\,\cos \theta } \right)^2+\left( {1-\beta ^2} \right)\,R^2\,\sin ^2\theta } \right]^{\frac{5}{2}}}} \right]
\]
Integration over the angles yields
\[
I=-e\,\beta ^2\,x\;\int\limits_0^\infty {R\,dR\,} \frac{\sqrt {\left( {R-x} \right)^2} \left( {R+x} \right)-\sqrt {\left( {R+x} \right)^2} \left( {R-x} \right)}{\sqrt {\left( {R-x} \right)^2} \,\sqrt {\left( {R+x} \right)^2} \,\left( {x^2-R^2\beta ^2} \right)^2}
\]
For $R>x$ the integrand vanishes, and for $R\le x$ the integral becomes
\[
I=-2\,e\,\beta ^2\int\limits_0^x {dR\,} \frac{x\,R}{\,\left( {x^2-R^2\beta ^2} \right)^2}=\frac{-\,e\,\beta ^2}{x\,\left( {1-\beta ^2} \right)}
\]

\vspace{.5cm}
\noindent \textbf{Acknowledgments}

\noindent It is a pleasure to thank Vladimir Onoochin for his extremely valuable contributions to the discussion of a very difficult subject. Peter Enders suggested introducing the potential $U$ in Sect. 2 which led to a substantial simplification of the analysis.

\end{large}

\vspace{.5cm}

\begin{thebibliography}{99}
\bibitem{Engelhardt1}{Engelhardt W 2005 Gauge invariance in classical electrodynamics \textit{Ann. Fond. Louis de Broglie }\textbf{30 }157-78}
\bibitem{Jackson}{J. D. Jackson, \textit{Classical Electrodynamics, }Second Edition, John Wiley {\&} Sons, Inc., 
New York (1975), Sect. 6.4}
\bibitem{Onoochin}{Onoochin V 2002 On non-equivalence of Lorentz and Coulomb gauges within classical electrodynamics \textit{Ann. Fond. Louis de Broglie }\textbf{27 }163-84}
\bibitem{Chubykalo}{Chubykalo A, Espinoza A, Alvarado Flores R 2011 Electromagnetic potentials without gauge transformations\textit{ Phys. Scr.} \textbf{84} 015009}
\bibitem{Engelhardt2}{Engelhardt W, Onoochin V 2012 Comment on 'Electromagnetic potentials without gauge transformations' \textit{Phys. Scr.} \textbf{85} 047001}
\bibitem{Engelhardt}{Engelhardt W 2012 On the solvability of Maxwell's equations \textit{Ann. Fond. Louis de Broglie }\textbf{37 }3-14}
\bibitem{Bartlett}{Bartlett D F, Corle T R 1985 Measuring Maxwell's Displacement Current Inside a Capacitor \textit{Phys. Rev. Lett}. \textbf{55 }59-62}
\end{thebibliography}
\end{document}